# Two Earth-sized planets orbiting Kepler-20

Francois Fressin[1], Guillermo Torres[1], Jason F. Rowe[2], David Charbonneau[1], Leslie A. Rogers[3], Sarah Ballard[1], Natalie M. Batalha[4], William J. Borucki[2], Stephen T. Bryson[2], Lars A. Buchhave[5,6], David R. Ciardi[7], Jean-Michel Désert[1], Courtney D. Dressing[1], Daniel C. Fabrycky[8], Eric B. Ford[9], Thomas N. Gautier III[10], Christopher E. Henze[2], Matthew J. Holman[1], Andrew Howard[11], Steve B. Howell[2], Jon M. Jenkins[12], David G. Koch[2], David W. Latham[1], Jack J. Lissauer[2], Geoffrey W. Marcy[11], Samuel N. Quinn[1], Darin Ragozzine[1], Dimitar D. Sasselov[1], Sara Seager[3], Thomas Barclay[2], Fergal Mullally[12], Shawn E. Seader[12], Martin Still[2], Joseph D. Twicken[12], Susan E. Thompson[12] & Kamal Uddin[13]

[1]Harvard-Smithsonian Center for Astrophysics, 60 Garden Street, Cambridge, Massachusetts 02138, USA.

[2]NASA Ames Research Center, Moffett Field, California 94035, USA.

[3]Department of Physics, Massachusetts Institute of Technology, Cambridge, Massachusetts 02139, USA.

[4]Dept. of Physics and Astronomy, San Jose State University, San Jose, California 95192, USA.

[5]Niels Bohr Institute, University of Copenhagen, DK-2100, Copenhagen, Denmark.

[6]Center for Star and Planet Formation, University of Copenhagen, DK-1350, Copenhagen, Denmark.

[7]NASA Exoplanet Science Institute/California Institute of Technology, Pasadena, California 91125, USA.

[8]Dept. of Astronomy and Astrophysics, University of California, Santa Cruz, California 95064, USA.

[9]Astronomy Dept., University of Florida, Gainesville, Florida 32111, USA.

[10]Jet Propulsion Laboratory/California Institute of Technology, Pasadena, California 91109, USA.

[11]Dept. of Astronomy, University of California, Berkeley, California 94720, USA.

[12]SETI Institute/NASA Ames Research Center, Moffett Field, California 94035, USA.

[13]Orbital Sciences Corporation/NASA Ames Research Center, Moffett Field, California 94035, USA.

**Since the discovery of the first extrasolar giant planets around Sun-like stars[1,2], evolving observational capabilities have brought us closer to the detection of true Earth analogues. The size of an exoplanet can be determined when it periodically passes in front of (transits) its parent star, causing a decrease in starlight proportional to its radius. The smallest exoplanet hitherto discovered[3] has a radius 1.42 times that of the Earth's radius ($R_\oplus$), and hence has 2.9 times its volume. Here we report the discovery of two planets, one Earth-sized ($1.03 R_\oplus$) and the other smaller than the Earth ($0.87 R_\oplus$), orbiting the star Kepler-20, which is already**

**known to host three other, larger, transiting planets**[4]**. The gravitational pull of the new planets on the parent star is too small to measure with current instrumentation. We apply a statistical method to show that the likelihood of the planetary interpretation of the transit signals is more than three orders of magnitude larger than that of the alternative hypothesis that the signals result from an eclipsing binary star. Theoretical considerations imply that these planets are rocky, with a composition of iron and silicate. The outer planet could have developed a thick water vapour atmosphere.**

Precise photometric time series gathered by the Kepler spacecraft[5] over eight observation quarters (670 days) have revealed five periodic transit-like signals in the G8 star Kepler-20, of which three have been previously reported as arising from planetary companions[4] (Kepler-20 b, Kepler-20 c and Kepler-20 d, with radii of $1.91 R_\oplus$, $3.07 R_\oplus$ and $2.75 R_\oplus$, and orbital periods of 3.7 days, 10.9 days and 77.6 days, respectively). The two, much smaller, signals described here recur with periods of 6.1 days (Kepler-20 e) and 19.6 days (Kepler-20 f) and exhibit flux decrements of 82 parts per million (p.p.m.) and 101 p.p.m. (Fig. 1), corresponding to planet sizes of $0.868^{+0.074}_{-0.096} R_\oplus$ (potentially smaller than the radius of Venus, $R_{\text{Venus}} = 0.95 R_\oplus$) and $1.03^{+0.10}_{-0.13} R_\oplus$. The properties of the star are listed in Table 1.

A background star falling within the same photometric aperture as the target and eclipsed by another star or by a planet produces a signal that, when diluted by the light of the target, may appear similar to the observed transits in both depth and shape. The Kepler-20 e and Kepler-20 f signals have undergone careful vetting to rule out certain false positives that might manifest themselves through different depths of odd- and even-numbered transit events, or displacements in the centre of light correlated with the flux variations[6]. High-spatial-resolution imaging shows no neighbouring stars capable of causing the signals[4]. Radial-velocity measurements based on spectroscopic observations with the Keck I telescope rule out stars or brown dwarfs orbiting the primary star, but they are not sensitive enough to detect the acceleration of the star due to these putative planetary companions[4].

To establish the planetary nature of these signals with confidence we must establish that the planet hypothesis is much more likely than that of a false positive. For this we used the BLENDER procedure[7–9], a technique used previously to validate the three smallest known exoplanets, Kepler-9 d (ref. [8]), Kepler-10 b (ref. [3]), and CoRoT-7 b (ref. [10]). The latter two were also independently confirmed with Doppler studies[3,11]. We used BLENDER to identify the allowed range of properties of blends that yield transit light curves matching the photometry of Kepler-20 e and Kepler-20 f. We varied as free parameters the brightness and spectral type (of the stars) or the size (for the planetary companions), the impact parameter, the eccentricity and the longitude of periastron. We simulated large numbers of these scenarios and compared the resulting light curves with the observations. We ruled out fits significantly worse (at the $3\sigma$ level, or greater) than that of a true transiting planet around the target, and we tabulated all remaining scenarios that were consistent with the Kepler light curves.

We assessed the frequency of blend scenarios through a Monte Carlo experiment in which we randomly drew $8 \times 10^5$ background main-sequence stars from a Galactic structure model[12] in a one-square-degree area around the target, and assigned them each a stellar or planetary transiting companion based on the known properties of eclipsing binaries[13] and the size distribution of planet candidates as determined from the Kepler mission itself[14]. We counted how many satisfy the constraints from BLENDER as well as observational constraints from our high-resolution imaging observations and centroid motion analysis[4], and made use of estimates of the frequencies of larger transiting planets and eclipsing binaries (see Fig. 2). In this way we estimated a blend frequency of background stars transited by larger planets of $2.1 \times 10^{-7}$ and a blend frequency of background eclipsing binaries of $3.1 \times 10^{-8}$, yielding a total of $2.4 \times 10^{-7}$ for Kepler-20 e. Similarly, $4.5 \times 10^{-7} + 1.26 \times 10^{-6}$ yields a total blend frequency of $1.7 \times 10^{-6}$ for Kepler-20 f.

Another type of false positive consists of a planet transiting another star physically associated with the target star. To assess their frequency we simulated $10^6$ such companions in randomly oriented orbits around the target, based on known distributions of periods, masses and eccentricities of binary stars[13]. We excluded those that would have been detected in our high-resolution imaging or that would have an overall colour inconsistent with the observed colour of the target, measured between the Sloan r band ($12.423 \pm 0.017$; ref. [14]) and the Warm Spitzer 4.5-µm band (10.85 ± 0.02; ref. 4). We used BLENDER to determine the range of permitted sizes for the planets as a function of stellar mass, and to each we assigned an eccentricity drawn from the known distribution for close-in exoplanets[15]. The frequency of blends of this kind is $5.0 \times 10^{-7}$ for Kepler-20 e, and $3.5 \times 10^{-6}$ for Kepler-20 f. Summing the contributions of background stars and physically bound stars, we find a total blend frequency of $7.4 \times 10^{-7}$ for Kepler-20 e and $5.2 \times 10^{-6}$ for Kepler-20 f.

We estimated the a priori chance that Kepler-20 has a planet of a similar size as implied by the signal using a $3\sigma$ criterion as in Blender, by calculating the fraction of Kepler objects of interest in the appropriate size range. We counted 102 planet candidates in the radius range allowed by the photometry of Kepler-20 e, and 228 for Kepler-20 f. We made the assumption that only 10% of them are planets (which is conservative in comparison to other estimates of the false positive rate that are an order of magnitude larger[16]). From numerical simulations, we determined the fraction of the 190,186 Kepler targets for which planets of the size of Kepler-20 e and Kepler-20 f could have been detected (17.4% and 16.0%, respectively), using actual noise levels. We then calculated the planet priors (a priori chance of a planet) to be $(102 \times 10\%)/(190,186 \times 17.4\%) = 3.1 \times 10^{-4}$ for Kepler-20 e, and $(228 \times 10\%)/(190,186 \times 16.0\%) = 7.5 \times 10^{-4}$ for Kepler-20 f. These priors ignore the fact that Kepler-20 is more likely to have a transiting planet at the periods of Kepler-20 e and Kepler-20 f than a random Kepler target, because the star is already known to have three other transiting planets, and multi-planet systems tend to be coplanar[17]. When accounting for this using the procedure described for the validation of Kepler-18 d (ref. 18), we find that the flatness of the system increases the transit probability from 7.7% to 63% for Kepler-20 e, and from 3.7% to 35% for Kepler-20 f. With this co-planarity boost, the planet priors increase to $2.5 \times 10^{-3}$ for Kepler-20 e and $7.1 \times 10^{-3}$ for Kepler-20 f. Comparing this with the total blend frequencies, we find that the hypothesis of an Earth-size planet for Kepler-20 e is 3,400 times more likely than that of a false positive, and 1,370 times for Kepler-20 f. Both of these odds ratios are sufficiently large to validate these objects with very high confidence as Earth-size exoplanets.

With measured radii close to that of the Earth, Kepler-20 e and Kepler-20 f could have bulk compositions similar to Earth's (approximately 32% iron core, 68% silicate mantle by mass; see Fig. 3), although in the absence of a measured mass the composition cannot be determined unambiguously. We infer that the two planets almost certainly do not have a hydrogen-dominated gas layer, because this would readily be lost to atmospheric escape owing to their small sizes and high equilibrium temperatures. A planet with several per cent water content by mass surrounding a rocky interior is a possibility for Kepler-20 f, but not for Kepler-20 e. If the planets formed beyond the snowline from a comet-like mix of primordial material and then migrated closer to the star, Kepler-20 f could retain its water reservoir for several billion years in its current orbit, but the more highly irradiated Kepler-20 e would probably lose its water reservoir to extreme-ultraviolet-driven escape within a few hundred million years[19]. In this scenario, Kepler-20 f could develop a thick vapour atmosphere with a mass of $0.05 M_\oplus$ that would protect the planet surface from further vaporization[20]. From the theoretical mass estimates in Table 1, we infer the semi-amplitude of the stellar radial velocity to be between 15 cm s$^{-1}$ and 62 cm s$^{-1}$ for Kepler-20 e and between 17 cm s$^{-1}$ and 77 cm s$^{-1}$ for Kepler-20 f. Such signals could potentially be detectable in the next few years, and would constrain the composition of the two planets.

**Acknowledgements** Kepler was competitively selected as the tenth Discovery mission. Funding for this mission is provided by NASA's Science Mission Directorate.

**Author Contributions** F.F. and G.T. developed the ideas and tools to perform the BLENDER analysis, and the statistical interpretation that led to the validation of Kepler-20 e and Kepler-20 f. C.E.H. implemented important modifications to the BLENDER program to improve the mapping of the range of possible blends. T.N.G. and D.C. led the effort to validate the largest planets in the Kepler-20 system. W.J.B. and D.G.K. led the Kepler mission, and supported the BLENDER effort on Kepler-20. N.M.B. led the Kepler science team that identified viable planet candidates coming out of the Kepler pipeline. J.F.R. and S.T.B. performed the light curve analysis to extract the planet characteristics. G.T., G.W.M., A.H., L.A.B., S.N.Q., D.W.L., D.C.F. and J.F.R. established the stellar characteristics from high-resolution spectroscopy and transit constraints. L.A.R., S.S. and D.D.S. worked on modelling the composition of the planets. J.M.J., T.B., F.M., S.E.S., M.S., S.E.T., J.D.T. and K.U. worked on the data collection, processing and review that yielded the time series photometry. D.R.C. provided the constraint on the angular separation from adaptive optics imaging using PHARO at Palomar. S.B.H. carried out speckle interferometry observations. S.T.B. worked on the pixel-level centroid analysis to reduce the sky area in which background binaries can reproduce the observed transits, which contributes to the statistical validation. D.C.F., E.B.F. and M.J.H. worked on the dynamical constraints on the planet properties. D.R. developed and calculated the coplanarity probability boost. C.D.D. worked on the Kepler incompleteness estimates. J.-M.D. analysed the Spitzer observations of Kepler-20. J.J.L. worked on estimating the planet prior. All authors discussed the results and commented on the manuscript. F.F. led the project and wrote the paper.

**Author Information** Reprints and permissions information is available at www.nature.com/reprints. The authors declare no competing financial interests. Readers are welcome to comment on the online version of this article at www.nature.com/nature. Correspondence and requests for materials should be addressed to F.F. (ffressin@cfa.harvard.edu).

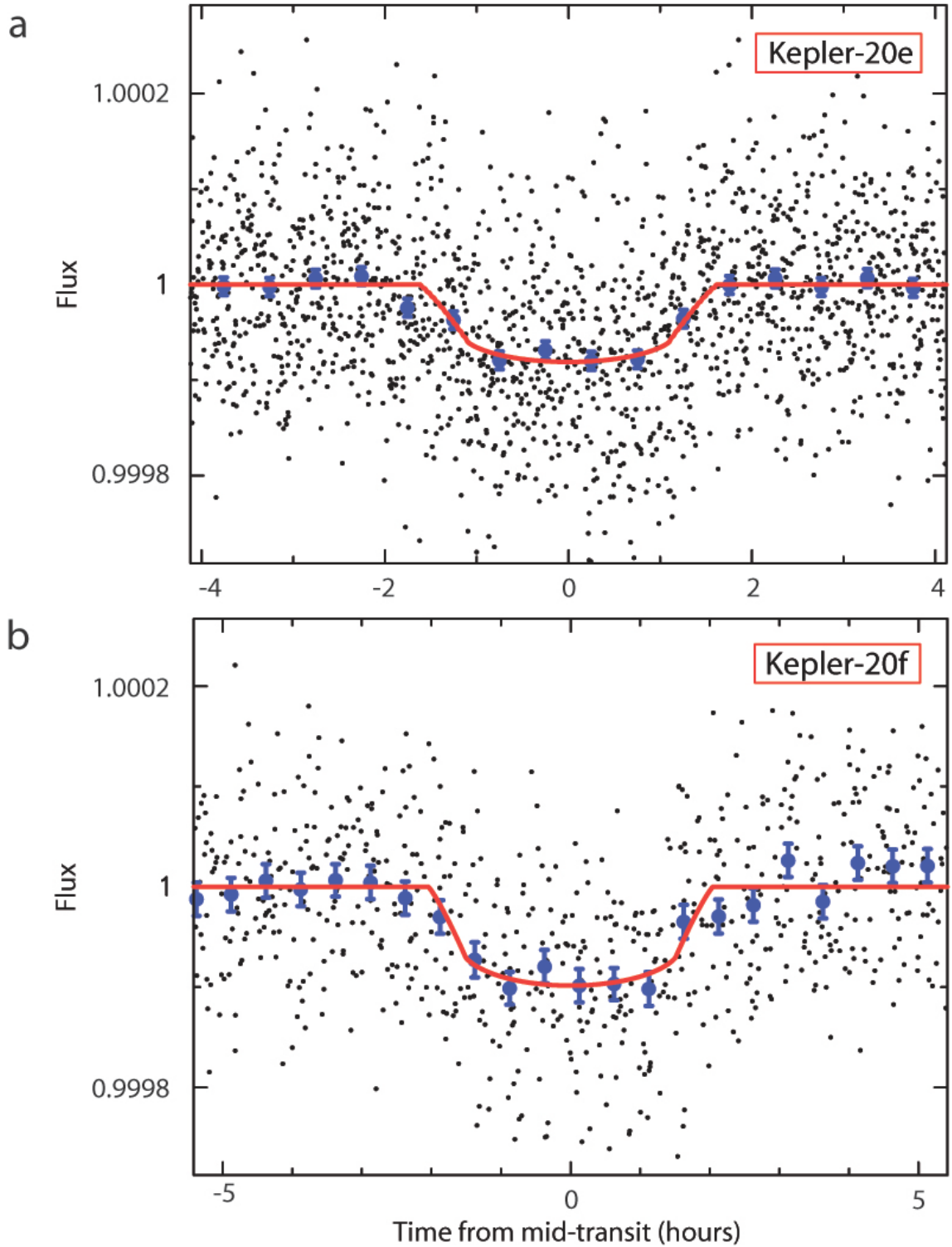

**Figure 1 Transit light curves.** Kepler-20 (also designated as KOI 070, KIC 6850504 and 2MASS J19104752+4220194) is a G8V star of Kepler magnitude 12.497 and celestial coordinates right ascension $\alpha = $ 19 h 10 min 47.5 s and declination $\delta = $ +42° 20′ 19.38″. The stellar properties are listed in Table 1. The photometric data

used for this work were gathered between 13 May 2009 and 14 March 2011 (quarter 1 to quarter 8), and comprise 29,595 measurements at a cadence of 29.426 min (black dots). The Kepler photometry phase-binned in 30-min intervals (blue dots with $1\sigma$ standard error of the mean (s.e.m.) error bars) for Kepler-20 e (**a**) and Kepler-20 f (**b**) is displayed as a function of time, with the data detrended[4] and phase-folded at the period of the two transits. Transit models (red curves) smoothed to the 29.426-min cadence are overplotted. These two signals are unambiguously detected in each of the eight quarters of Kepler data, and have respective signal-to-noise ratios of 23.6 and 18.5, which cannot be due to stellar variability, data treatment or aliases from the other transit signals[4].

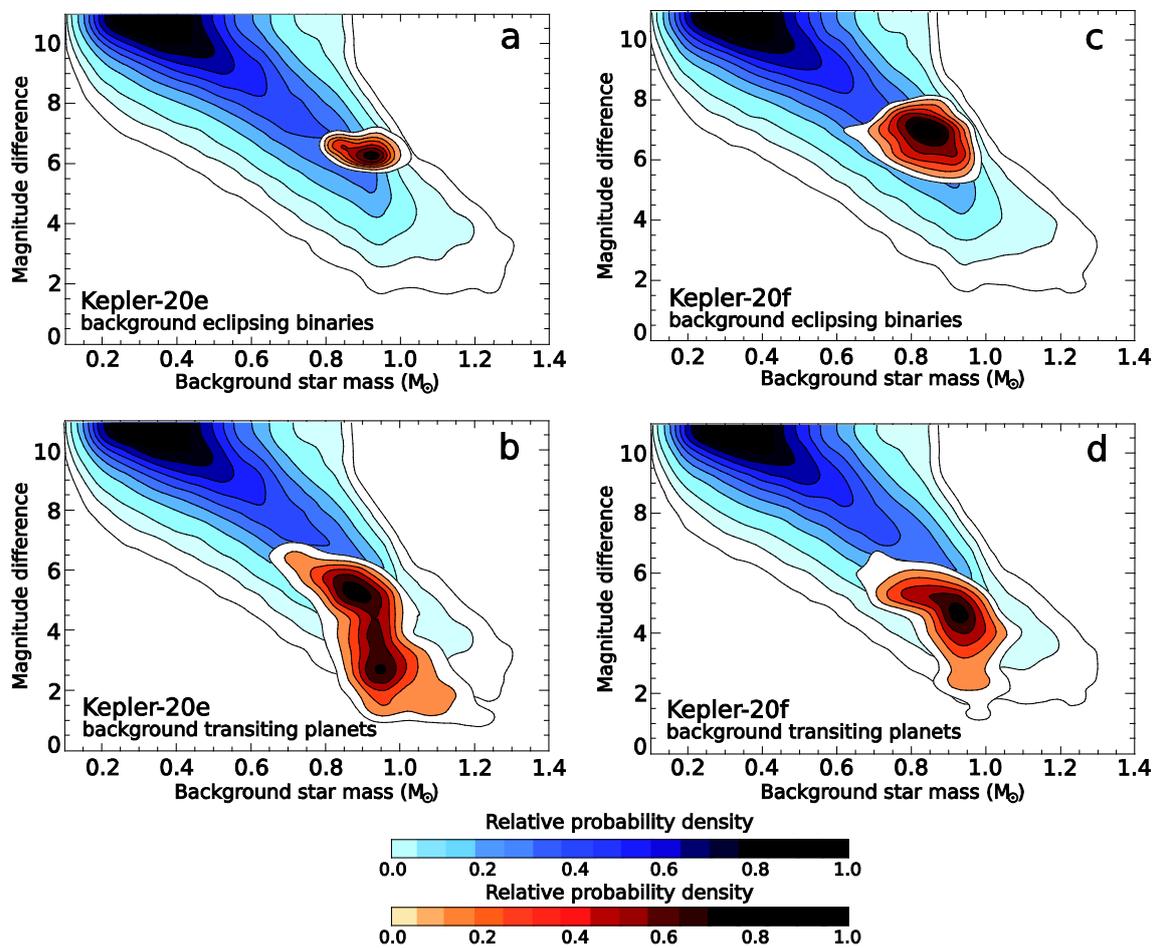

**Figure 2 Density map of stars in the background of Kepler-20.** The blue-shaded contours correspond to main-sequence star counts from the Besancon model in the vicinity of Kepler-20, as a function of stellar mass and magnitude difference in the Kepler passband compared to Kepler-20. The red-shaded contours represent the fractions of those stars orbited by another smaller star (**a** and **c**) or by a planet (**b** and **d**)

with sizes such that the resulting light curves mimic the transit signals for Kepler-20 e and Kepler-20 f. The displacement of the blue and red contours in magnitude and spectral type results in very small fractions of the simulated background stars being viable false positives for Kepler-20 e (1.6% when transited by a planet, and 0.1% when transited by a smaller star). We obtained similar results for Kepler-20 f (2.1% when transited by a planet, and 3.1% when transited by a smaller star). Most of these background stars have masses (spectral types) near that of the target, and are two to seven magnitudes fainter. The above fractions are further reduced because background stars able to match the signals but that are bright enough and at large enough angular separation from the target would have been detected in our imaging observations and/or centroid motion analysis. Finally, to obtain the blend frequencies we scaled these estimates to account for the fraction of background stars expected to have transiting planets (1.29%, the ratio between the number of Kepler objects of interest and the total number of Kepler targets[25]) or stellar companions (0.79% based on the statistics of detached eclipsing binaries in the Kepler field[26]). We examined non-main-sequence stars as alternatives to either object of the blend eclipsing pair, but found that they either do not reproduce the observed transit shape well enough, or are much less common (<1%) than main-sequence blends.

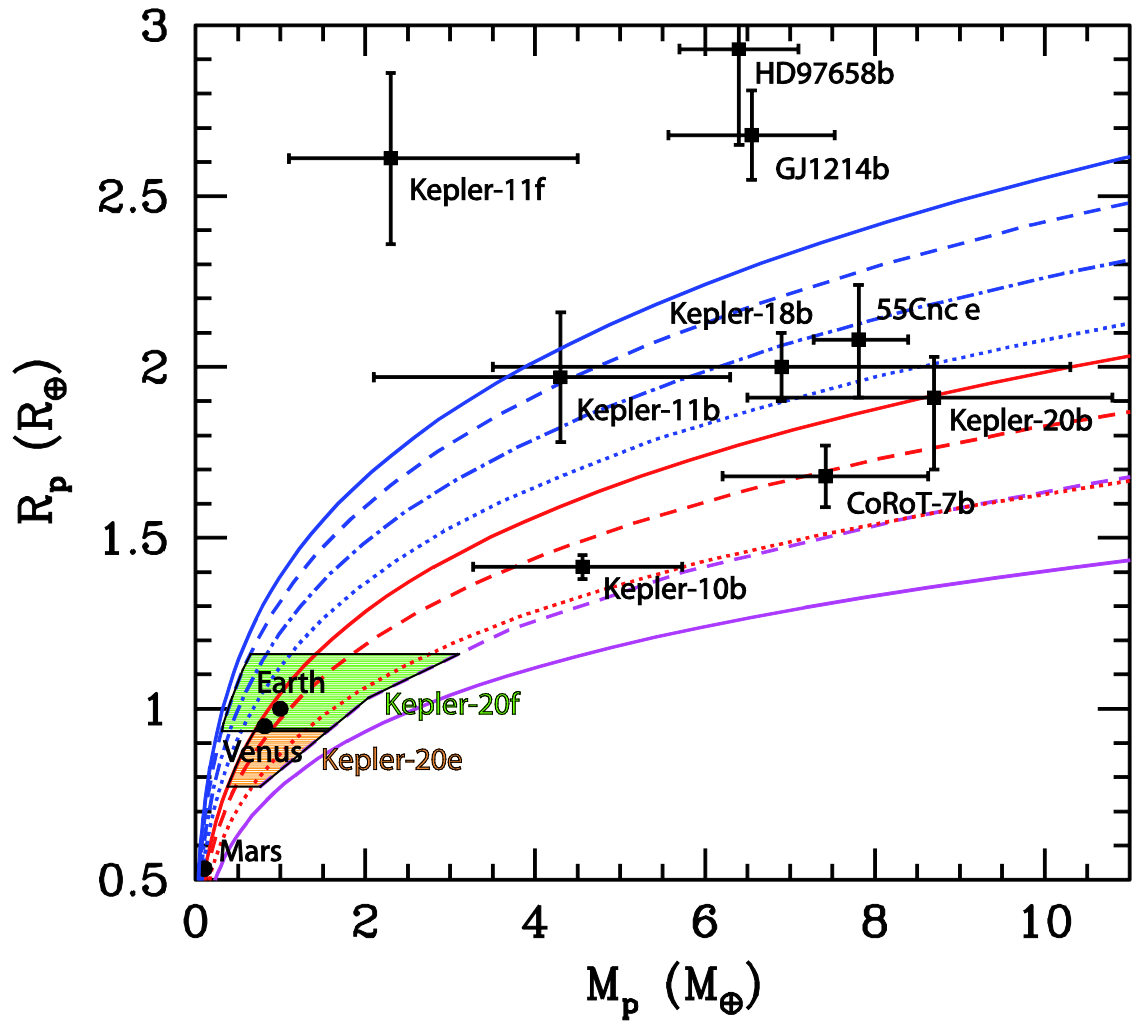

**Figure 3 Mass versus radius relation for small planets.** Kepler-20 e and Kepler-20 f theoretical mass and observed radius ranges ($1\sigma$) are plotted as orange- and green-shaded areas, while the other transiting planets with dynamically determined masses are plotted in black, with $1\sigma$ error bars. The curves are theoretical constant-temperature mass–radius relations[27]. The solid lines are homogeneous compositions: water ice (solid blue), $MgSiO_3$ perovskite (solid red), and iron (magenta). The non-solid lines are mass–radius relations for differentiated planets: 75% water ice, 22% silicate shell and 3% iron core (dashed blue); Ganymede-like with 45% water ice, 48.5% silicate shell and 6.5% iron core (dot-dashed blue); 25% water ice, 52.5% silicate shell and 22.5% iron core (dotted blue); approximately Earth-like with 67.5% silicate mantle and 32.5% iron core (dashed red); and Mercury-like with 30% silicate mantle and 70% iron core (dotted red). The dashed magenta curve corresponds to the density limit from a formation model[24].

The minimum density for Kepler-20 e corresponds to a 100% silicate composition, because this highly irradiated small planet could not keep a water reservoir. The minimum density for Kepler-20 f follows the 75% water-ice composition, representative of the maximum water content of comet-like mix of primordial material in the solar system[28].

## Table 1. Stellar and planetary parameters for Kepler-20.

*Stellar properties*

| | |
|---|---|
| Effective temperature, $T_{eff}$ | 5,466 ± 93 K |
| Surface gravity log g (cm s$^{-2}$) | 4.443 ± 0.075 |
| Metallicity [Fe/H] | 0.02 ± 0.04 |
| Projected rotational velocity, v sin i | 0.4 ± 0.5 km s$^{-1}$ |
| Stellar mass, $M_s$ | 0.912 ± 0.035 $M_\odot$ |
| Stellar radius, $R_s$ | 0.944 +0.060,-0.095 $R_\odot$ |
| Stellar density, $\rho_s$ | 1.51 ± 0.39 g cm$^{-3}$ |
| Luminosity, $L_s$ | 0.853 ± 0.093 $L_\odot$ |
| Distance, D | 290 ± 30 pc |

*Transit parameters and physical properties: Kepler-20e (KOI-070.04)*

| | |
|---|---|
| Orbital period, P | 6.098493 ± 0.000065 days |
| Time of centre of transit, $T_c$ | 2,454,968.9336 ± 0.0039 BJD |
| Eccentricity, e | < 0.28 |
| Planet/star radius ratio, $R_p/R_s$ | 0.00841 +0.00035,-0.00054 |
| Scaled semi-major axis, $a/R_s$ | 11.56 +0.21,-0.29 |
| Impact parameter, b | 0.630 +0.070,-0.053 |
| Orbital inclination, i | 87.50 +0.33,-0.34 degrees |
| Planetary radius, $R_p$ | 0.868 +0.074,-0.096 $R_\oplus$ |

| Planetary mass, $M_p$ | < 3.08 $M_\oplus$ (spectroscopic limit) |
| --- | --- |
| | 0.39 $M_\oplus$ < $M_p$ < 1.67 $M_\oplus$ (theoretical considerations) |
| Planetary equilibrium temperature, $T_{eq}$ | 1040 ± 22 K |

*Transit parameters and physical properties: Kepler-20f (KOI-070.05)*

| Orbital period, P | 19.57706 ± 0.00052 days |
| --- | --- |
| Time of centre of transit, $T_c$ | 2,454,968.219 ± 0.011 BJD |
| Eccentricity | < 0.32 |
| Planet/star radius ratio, $R_p/R_s$ | 0.01002 +0.00063,-0.00077 |
| Scaled semi-major axis, $a/R_s$ | 25.15 +0.47,-0.63 |
| Impact parameter, b | 0.727 +0.054,-0.053 |
| Orbital inclination, i | 88.68 +0.14,-0.17 degrees |
| Planetary radius, $R_p$ | 1.03 +0.10,-0.13 $R_\oplus$ |
| Planetary mass, $M_p$ | < 14.3 $M_\oplus$ (spectroscopic limit) |
| | 0.66 $M_\oplus$ < $M_p$ < 3.04 $M_\oplus$ (theoretical considerations) |
| Planetary equilibrium temperature, $T_{eq}$ | 705 ± 16 K |

$M_\odot$, mass of the Sun; $R_\odot$, radius of the Sun. The effective temperature, surface gravity, metallicity and projected rotational velocity of the star were spectroscopically determined[21] from our Keck/HIRES spectrum. With these values and the use of stellar evolution models[22], we derived the stellar mass, radius, luminosity, distance and mean density. The transit and orbital parameters (period, time of centre of transit, radius ratio, scaled semi-major axis, impact parameter and orbital inclination) for the five planets in the Kepler-20 system were derived jointly based on the Kepler photometry using a Markov-chain Monte Carlo procedure with the mean stellar density as a prior[4]. The parameters above are based on an eccentricity constraint: that the orbits do not cross each other. After calculating the above parameters, we performed a suite of *N*-body integrations to estimate the maximum eccentricity for each planet consistent with dynamical stability[4]. The *N*-body simulations provide similar constraints on the maximum eccentricity and justify the assumption of non-crossing orbits. The planetary spectroscopic mass limits are the $2\sigma$ upper limits determined from the radial velocity analysis based on the Keck radial velocity measurements. Planet interior models provide further useful constraints on mass and inferences on composition[23]. Assuming Kepler-20 e and Kepler-20 f are rocky bodies comprised of iron and silicates, and considering the uncertainty on their radii, the planet masses are constrained to be $0.39 M_\oplus < M_p < 1.67 M_\oplus$ for Kepler-20 e, and $0.66 M_\oplus < M_p < 3.04 M_\oplus$ for Kepler-20 f. The lower and upper mass bounds are set by a homogeneous silicate composition and by the densest composition from a model of planet formation with collisional mantle stripping[24]. The planet equilibrium temperatures assume an Earth-like Bond albedo of 0.3, isotropic redistribution of heat for reradiation, and a circular orbit. The errors in these quantities reflect only the uncertainty due to the stellar luminosity.